\documentclass[preprint,onecolumn,nofootinbib]{revtex4}
\usepackage{hyperref,graphicx}
\pdfoutput=1
\usepackage{epsfig}
\usepackage{amsmath}
\usepackage{amsfonts}
\usepackage{amssymb}
\usepackage{color}%
\usepackage{dcolumn}
\providecommand{\U}[1]{\protect\rule{.1in}{.1in}}
\newcommand{\f}{\begin{equation}}
\newcommand{\ff}{\end{equation}}
\newcommand{\fa}{\begin{eqnarray}}
\newcommand{\ffa}{\end{eqnarray}}
\begin{document}
\title{Holographic Superconductor on Q-lattice}
\author{Yi Ling $^{1,2}$}
\email{lingy@ihep.ac.cn}
\author{Peng Liu $^{1}$}
\email{liup51@ihep.ac.cn}
\author{Chao Niu $^{1}$}
\email{niuc@ihep.ac.cn}
\author{Jian-Pin Wu $^{3,2}$}
\email{jianpinwu@gmail.com}
\author{Zhuo-Yu Xian $^{1}$}
\email{xianzy@ihep.ac.cn}
\affiliation{$^1$ Institute of High
Energy Physics, Chinese
Academy of Sciences, Beijing 100049, China\\
$^2$ State Key Laboratory of Theoretical Physics, Institute of
Theoretical Physics, Chinese Academy of Sciences, Beijing 100190,
China\\
$^3$ Department of Physics, School of Mathematics and Physics,
Bohai University, Jinzhou 121013, China}
\begin{abstract}
We construct the simplest gravitational dual model of a
superconductor on Q-lattices. We analyze the condition for the
existence of a critical temperature at which the charged scalar
field will condense. In contrast to the holographic superconductor
on ionic lattices, the presence of Q-lattices will suppress the
condensate of the scalar field and lower
the critical temperature.
 In particular, when the Q-lattice background is dual to a
deep insulating phase, the condensation would never occur for some
small charges. Furthermore, we numerically compute the optical
conductivity in the superconducting regime. It turns out that the
presence of Q-lattice does not remove the pole in the imaginary
part of the conductivity, ensuring the appearance of a delta
function in the real part. We also evaluate the gap which in
general depends on the charge of the scalar field as well as the
Q-lattice parameters. Nevertheless, when the charge of the scalar
field is relatively large and approaches the probe limit, the gap
becomes universal with $\omega_g\simeq 9T_c$ which is consistent
with the result for conventional holographic superconductors.
\end{abstract}
\maketitle

\section{Introduction}
The Gauge/Gravity duality has provided a powerful tool to
investigate many important phenomena of strongly correlated system
in condensed matter physics. One remarkable achievement is the
building of a gravitational dual model for a
superconductor\cite{Gubser:2008px,Hartnoll:2008vx,Hartnoll:2008kx}.
This construction has been extensively investigated in literature
and more and more evidences in favor of this approach have been
accumulated. In particular, the holographic lattice technique
proposed recently has brought this approach into a new stage to
reproduce quantitative features of realistic
 materials in
experiments\cite{Horowitz:2012ky,Horowitz:2012gs,Horowitz:2013jaa}.
One remarkable achievement in this direction is {the successful description of} the Drude behavior of the optical
conductivity at low frequency
regime\cite{Horowitz:2012ky,Horowitz:2012gs,Horowitz:2013jaa,
Hartnoll:2012PRL,Maeda:2012To,CSPark:2010Mo,Karch:2009La,Karch:2009Di,Sachdev:2013Cr,Mozaffar:2013Cr,
Wong:2013Cr,Chesler:2013qla,Vegh:2013Ma,Davison:2013Ma,Tong:2013Ma,Ling:2013nxa,Withers:2013loa,Andrade:2013gsa,Gouteraux:2014hca,Lucas:2014zea,Zeng:2014uoa}
and {the exhibition of} a band structure with Brillouin
zones\cite{Liu:2012tr,Ling:2013aya}. Inspired by
 holographic lattice techniques people have also developed numerical methods to
 construct spatially modulated phases with a spontaneous breaking of the translational invariance
 \cite{Ooguri:2010xs,Donos:2011bh,DonosHartnoll,Rozali:2012es,Rozali:2013ama,Donos:2013gda,Donos:2013wia,Ling:2014saa,Jokela:2014dba}.

The original holographic lattice with full backreactions is
simulated by a real scalar field or chemical potential which has a
periodic structure on the boundary of space
time\cite{Horowitz:2012ky,Horowitz:2012gs,Horowitz:2013jaa}. We
may call these lattices as scalar lattice and ionic lattice,
respectively. This framework contains one limitation during the
course of application. Namely, the numerical analysis involves
a group of partial differential equations to solve, while its accuracy heavily depends on the temperature of the background
which usually are black hole ripples.
Thus, it is very challenging
to explore the lattice effects at very low or even zero
temperature(for recent progress, see \cite{Hartnoll:2014gaa}).
Very recently, another much simpler but elegant framework
for
constructing holographic lattices is proposed in
\cite{Donos:2013eha}, which is dubbed as the Q-lattice, because of
some analogies with the construction of
Q-balls\cite{Coleman:1985ki}\footnote{Another sort of simpler
holographic models with momentum relaxation can be found in
\cite{Andrade:2013gsa,Gouteraux:2014hca}, where a family of black
hole solutions is
 characterized only by $T$ and $k$, while the parameter
representing the lattice amplitude in Q-lattice is absent, thus no
metal-insulator transition at low temperature.}. In this
framework, one only need
 to solve the ordinary differential
equations to compute the transport coefficients of the system,
thus numerically one may drop the temperature down to a regime
which may exhibit some new physics. Indeed, one novel feature has
been observed in this framework. It is disclosed that black hole
solutions at a fairy low temperature may be dual to different
phases and a metal-insulator transition can be implemented by
adjusting the parameters of
Q-lattices\cite{Donos:2013eha,Donos:2014uba,Donos:2014oha,Donos:2014yya}.
Another advantage of Q-lattice framework is that the charge
density as well as the chemical potential on the boundary can
still be uniformly distributed even in the presence of the lattice
background, which seems to be closer to a practical lattice system
in condensed matter physics. However,
 in the context of ionic
lattice the presence of the lattice structure always brings out a
periodically distributed charge density and chemical potential on
the boundary, which looks peculiar from the side of the condensed
matter physics.

In this paper we intend to investigate the Q-lattice effects on
holographic superconductor models.
 We will show that in general the presence of Q-lattices
will suppress the condensation of the scalar field and lower the critical temperature,
 which is in contrast to the
holographic superconductor on ionic lattices, where the critical
temperature is usually enhanced by the lattice
effects\cite{Ganguli:2012up,Horowitz:2013jaa}. In particular, when
the black hole background is dual to a deep insulating phase, the
condensation would never occur for some small charges.
Furthermore, we will numerically compute the optical conductivity
in the superconducting regime. It turns out that the Q-lattice
does not remove the pole in the imaginary part of the
conductivity, implying the appearance of a delta function in the
real part and ensuring that the superconductivity is genuine and
not due to the translational invariance. We also evaluate the gap
with a result $\omega_g\simeq 9T_c$ in the probe limit, which is
almost independent of the parameters of Q-lattices.

We organize the paper as follows. In next section we present the
holographic setup for the superconductor model on Q-lattice, and
briefly review the black hole backgrounds which are dual to
metallic phases and insulating phases, respectively. Then in
section three we will analyze the instability of these solutions
and numerically compute the critical temperature for the
condensate of the charged scalar field. The optical conductivity
in the direction of the lattice will be given in section four and
the gap will be evaluated as well. We conclude with some comments
in section five.

\section{The holographic setup}\label{Setup}
{Recently} various investigations {to}
 the inhomogeneous effects or
lattice effects on
  holographic superconductors have been presented in literature\cite{Flauger:2010tv,Hutasoit:2011rd,Ganguli:2012up,Hutasoit:2012ib,Alsup:2012kr,Maeda:2012To,Erdmenger:2013zaa,Kuang:2013jma,Arean:2013mta,Zeng:2013yoa},
but these
 effects are almost treated perturbatively and the
full backreaction on the metric is ignored. As far as we know, the
first lattice model of a holographic superconductor with full
backreaction is constructed in \cite{Horowitz:2013jaa}, in a
framework of ionic lattice. Here, inspired by the recent work on
 Q-lattices in\cite{Donos:2013eha}, we will construct an alternative
lattice model of holographic superconductor closely following the
route presented in \cite{Horowitz:2013jaa}. As the first step we
will construct the simplest model with the essential ingredients
in this paper, but leave all the other possible constructions for
further investigation in future. We start from a gravity model
with two complex scalar fields plus a $U(1)$ gauge field in four
dimensions. If we work in unit in which the $AdS$ length scale
$L=1$, then action is
\begin{eqnarray}
S&=&\frac{1}{2\kappa^2}\int
d^4x\sqrt{-g}[R+6-\frac{1}{2}F^{\mu\nu}F_{\mu\nu}+2\Psi\Psi^{*}-|(\nabla-ieA)\Psi|^2-|\nabla
\Phi|^2-m^2|\Phi|^2], \label{eq:action}
\end{eqnarray}
where $\Phi$ is neutral with respect to the Maxwell field and will
be responsible for the breaking of the translational invariance
and the formation of a Q-lattice background, while $\Psi$ is
charged under the Maxwell field and will be responsible for the
spontaneous breaking of the $U(1)$ gauge symmetry and the
formation of a superconducting phase.  For convenience, we may
further rewrite the $U(1)$ charged complex scalar field $\Psi$ as
a real scalar field $\eta$ and a St\"{u}ckelberg field
$\theta$,namely $\Psi=\eta e^{i\theta}$, such that the action
reads as
\begin{eqnarray}
S&=&\frac{1}{2\kappa^2}\int
d^4x\sqrt{-g}[R+6-\frac{1}{2}F^{\mu\nu}F_{\mu\nu}+(2-e^2A_{\mu}A^{\mu})\eta^2-(\partial
\eta)^2-|\partial \Phi|^2-m^2|\Phi|^2],\label{eq:action}
\end{eqnarray}
where we have fixed the gauge $\theta=0$. The equations of motion
can be obtained as
\begin{eqnarray}
&&R_{\mu\nu}-g_{\mu\nu}(-3+\frac{m^2}{2}|\Phi|^2-\eta^2)-\partial_{(\mu}\Phi\partial_{\nu)}\Phi^*-(F_{\mu\lambda}{F_{\nu}{}^{\lambda}}-\frac{1}{4}g_{\mu\nu}F^2)-
\partial_{\mu}\eta\partial_{\nu}\eta-e^2\eta^2A_{\mu}A_{\nu}=0,\nonumber\\
&&\nabla_{\mu}{F^{\mu}}_{\nu}-e^2\eta^2A_{\nu}=0,\nonumber\\
&&(\nabla^2-m^2)\Phi=0,\ \ \ \ \ \ \ \ \
(\nabla^2-e^2A^2+2)\eta=0.\label{eqom}
\end{eqnarray}

Obviously, in the case of $\eta=0$, the equations of motion can
give rise to the electric Reissner-Nordstr\"{o}m-AdS (RN-AdS) black hole solutions on Q-lattice
which have been constructed in \cite{Donos:2013eha}. This is a
three-parameter family of black holes characterized by the
temperature $T/\mu$,the lattice amplitude $\lambda/\mu^{3-\Delta}$
and the wave vector $k/\mu$, where $\mu$ is the chemical potential
of the dual field theory and can be treated as the unit for the
grand canonical system. The ansatz for the Q-lattice background is
\fa ds^2&=&{1\over
z^2}\left[-(1-z)p(z)Udt^2+\frac{dz^2}{(1-z)p(z)U}+V_1dx^2+V_2dy^2\right],\nonumber\\
A&=&\mu(1-z)\psi dt,\nonumber\\
\Phi &=& e^{ikx}z^{3-\Delta}\phi,\ffa with
$p(z)=1+z+z^2-\mu^2z^3/2$ and $\Delta=3/2\pm(9/4+m^2)^{1/2}$.
Notice that $U,V_1,V_2,\psi$ and $\phi$ are functions of the
radial coordinate $z$ only. Obviously if we set $U=V_1=V_2=\psi=1$
and $\phi=0$, the solution goes back to the familiar
RN-AdS metric. The non-trivial Q-lattice
solutions can be obtained by setting a non-trivial boundary
condition at infinity for the scalar field $\phi(0)=\lambda$ and
regular boundary conditions on the horizon, which is located at
$z=1$. The Hawking temperature of the black hole is $
T/\mu=(6-\mu^2)U(1)/(8\pi\mu)$. Through this paper we will fix
$m^2=-3/2$ such that the $AdS_2$ BF bound will not be violated.

It is shown in \cite{Donos:2013eha} that at low temperature the
system exhibits both metallic and insulating phases. In metallic
phase the conductivity in the low frequency regime is subject to
the Drude law and the DC conductivity climbs up with the decrease
of the temperature, while in insulating phase one can observe a
soft gap in the optical conductivity and the DC conductivity goes
down with the temperature. Numerically, one finds a small
lattice amplitude $\lambda$ will corresponds to a metallic phase
while a large one will be an insulating phase. In next section we
will discuss the condensate of the charged scalar field over
Q-lattice background.

\section{Background}
In this section we construct a Q-lattice background with a
condensate of the charged scalar field, namely $\eta\neq 0$.
Firstly, we intend to
justify  when the Q-lattice background
becomes unstable by estimating the critical temperature for
the formation of
 charged scalar hair. For this purpose we may treat the
equation of motion  of
 $\eta$ perturbatively. Namely, we intend to
find static normalizable mode of charged scalar field on a fixed
Q-lattice background. As argued in \cite{Horowitz:2013jaa}, it is
more convenient to turn this problem into a positive self-adjoint
eigenvalue problem for $e^2$, thus we rewrite the equation of
motion as the following form
\begin{equation}
-(\nabla^2+2)\eta=-e^2A^2\eta.\label{eqoms}
\end{equation}

\begin{figure}
\center{
\includegraphics[scale=1.2]{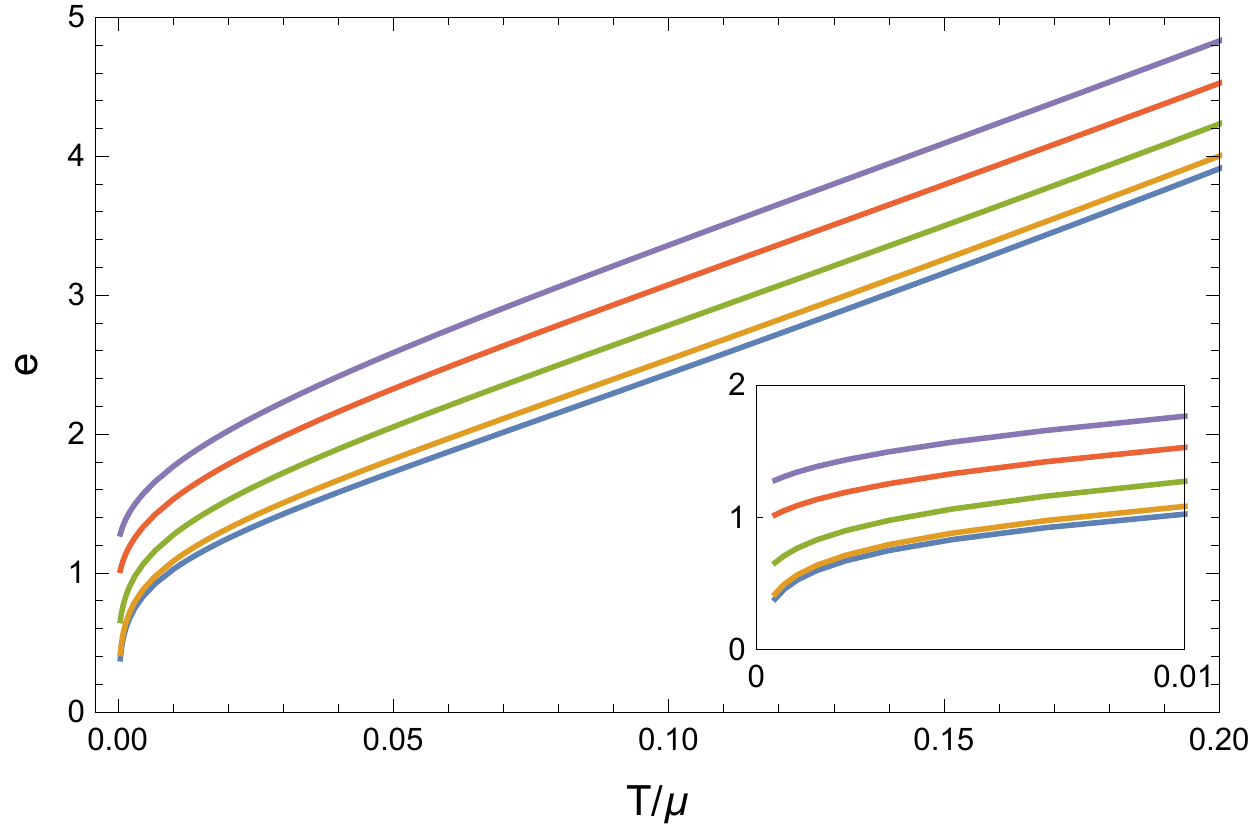}\hspace{0.5cm}
\caption{\label{solu1}The charge of the scalar field versus the
critical temperature for several values of the lattice amplitude.
From top to down the lines represent
$\lambda/\mu^{3-\Delta}=2,3/2,1,1/2,0$, where we have set
$k/\mu=1/\sqrt{2}$. } }
\end{figure}

\begin{figure}
\center{
\includegraphics[scale=0.38]{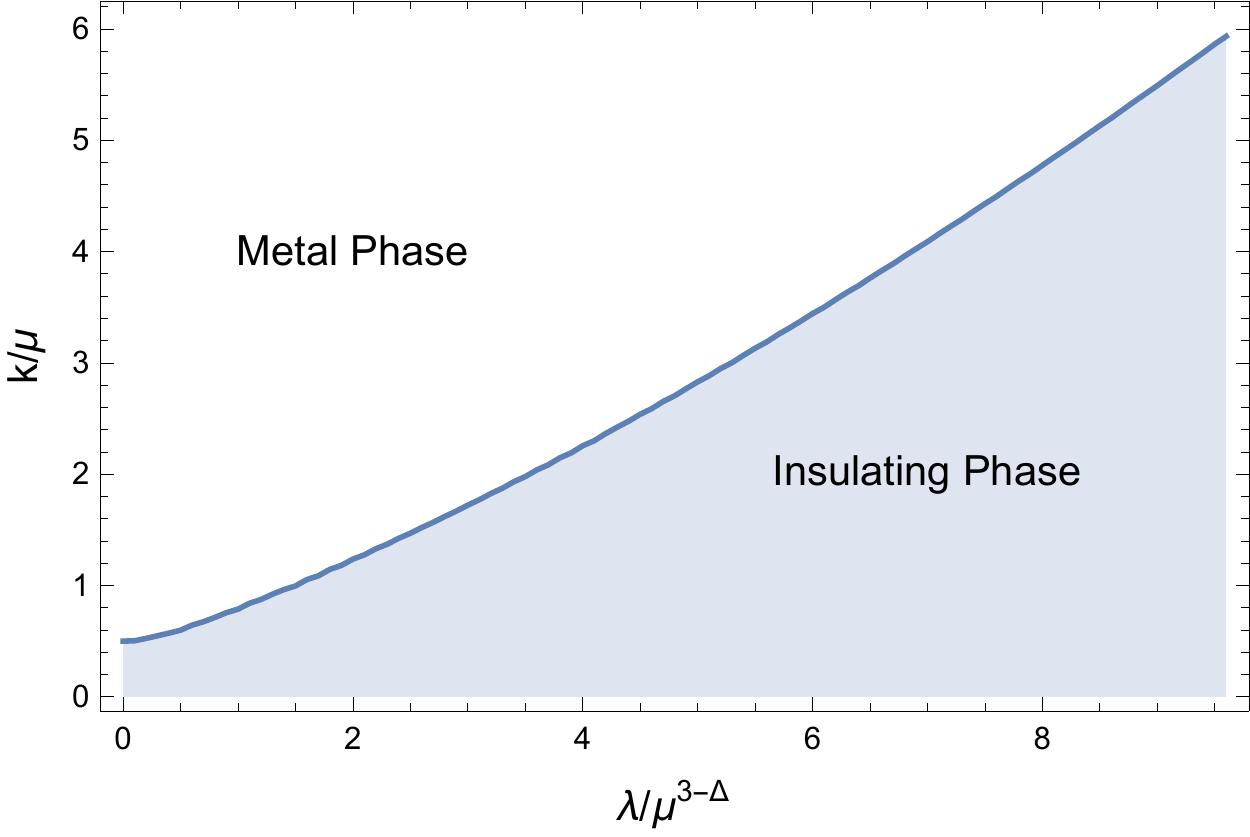}\hspace{0.1cm}
\includegraphics[scale=0.42]{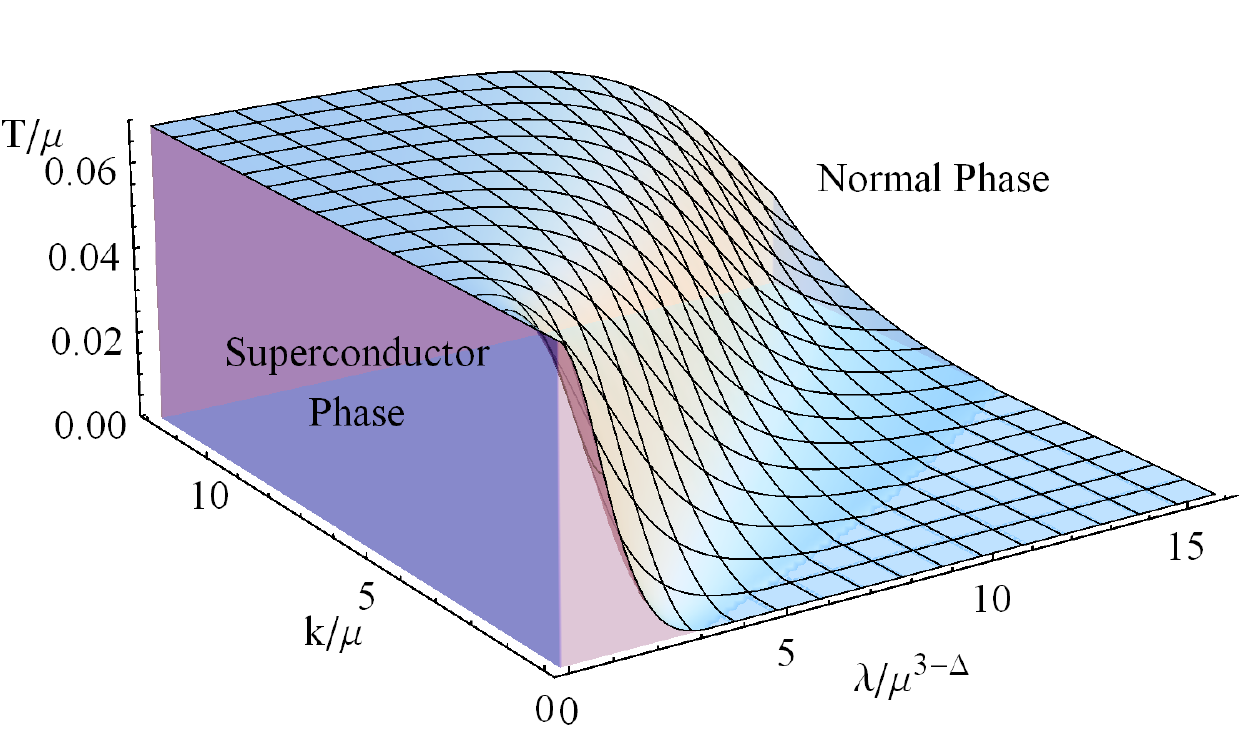}\hspace{0.5cm}
\includegraphics[scale=0.4]{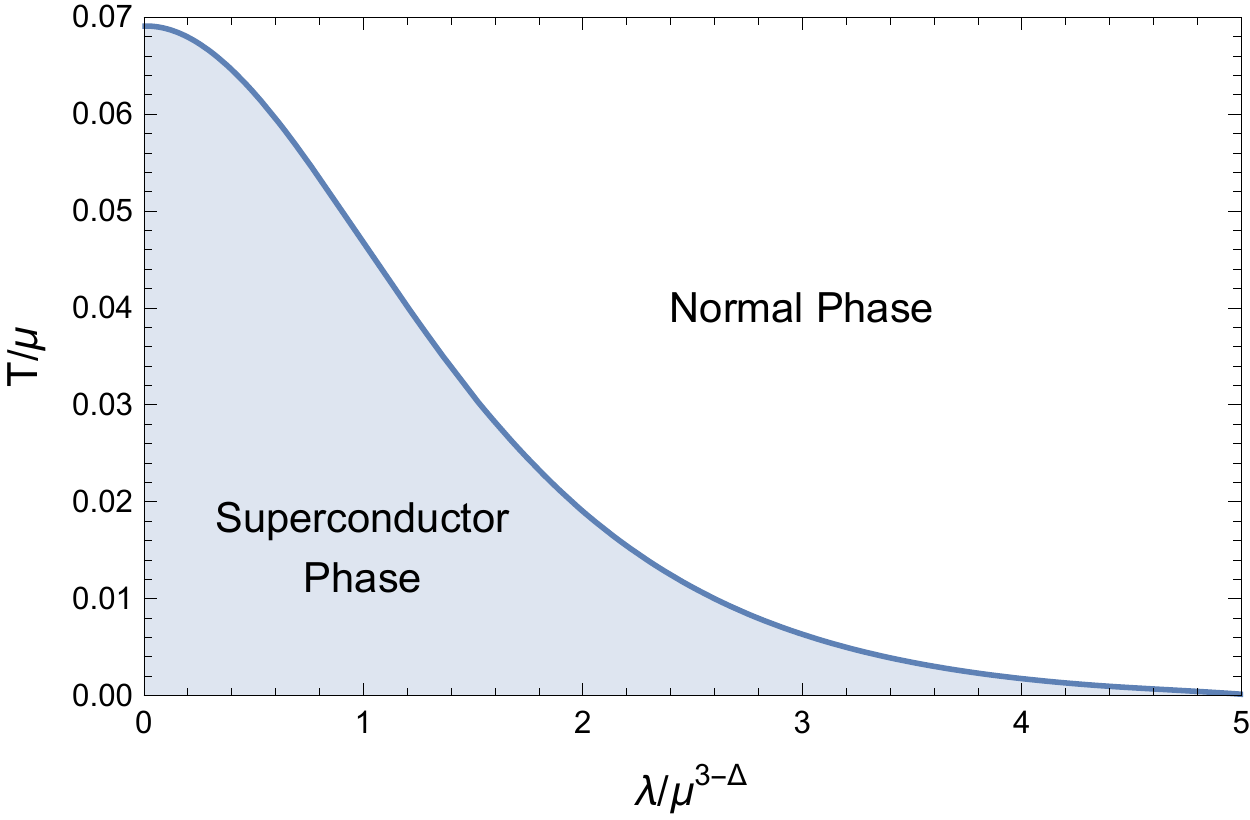}\hspace{0.1cm}
\caption{\label{solu2}The left plot is a phase diagram for pure
Q-lattice system without superconducting setup at extremely small
temperature $T/\mu=0.001$. The middle plot is a 3D plot for the
critical temperature as a function of the lattice parameters, with
$e=2$. Right plot is a phase diagram for a given wave number
$k/\mu=1/\sqrt{2}$. } }
\end{figure}

Before solving this equation numerically, we briefly discuss the
boundary condition for $\eta$. Without loss of generality, we have
set the mass of the charged scalar field as $m_\eta^2=-2$ from the
beginning, such that its asymptotical behavior at infinity is
\begin{equation}
\eta=z\eta_1+z^2\eta_2+...\label{dec}.
\end{equation}
In the dual theory $\eta_1$ is treated as the source and $\eta_2$
as the expectation value. Since we expect the condensate will turn
on without being sourced, we set $\eta_1=0$ through this paper.
Now imposing the regularity condition on the horizon and requiring
the scalar field to decay as in (\ref{dec}), one can find the
critical temperature for the condensate of the scalar field  by
solving the eigenvalue equation (\ref{eqoms}) for different values
of the charge $e$. Our results are shown in Fig.\ref{solu1}. There are
several curves on this plot, and each of them
{denotes}
 the change
of charge with the critical temperature for a fixed lattice
amplitude $\lambda$. It is very interesting to compare our results
here with those obtained in the ionic lattice
model\cite{Horowitz:2013jaa}.
\begin{itemize}
    \item As expected, each curve shows a rise in critical
    temperature with charge, which is consistent with our
    intuition that the increase of the charge make the
    condensation easier, thus the critical temperature becomes
    higher. This tendency is the same as that in \cite{Horowitz:2013jaa}.
    \item For a given charge,
     we find that increasing the lattice amplitude lowers the critical temperature, which means that the condensate of the scalar field is suppressed
    by the presence of the Q-lattice. Later we will find this tendency can be further
    confirmed by plotting the value of the condensate as a function of
    temperature. Such a tendency is contrary to what have been found in ionic lattice and striped
    superconductors, where the critical temperature is enhanced by
the lattice effects\cite{Ganguli:2012up,Horowitz:2013jaa}.
Preliminarily we think
 this discrepancy
    might come from the different behaviors of the chemical
    potential in different lattice backgrounds, as analyzed in \cite{Horowitz:2013jaa}, where $\mu$ is manifestly periodic
    while in Q-lattice model all the fields does not manifestly depend on $x$
    except the scalar field $\Phi$.
    \item At the zero temperature limit, not all the curves
    have a tendency to
    converge to the same point as depicted in ionic lattices\cite{Horowitz:2013jaa}.
    On the contrary, we find when the amplitude of the lattice is
    large enough (which may correspond to an insulating phase before the occurrence of the condensation),
    these lines do not converge at least at the temperature regime that our numerical accuracy can reach (See the inset of Fig.\ref{solu1}).
    It implies that for a given charge if its value is relatively small, the system with large lattice amplitude
    would not undergo a phase transition
    no matter how low the
    temperature is! For instance, if $e=1$, there would
    be no phase transition for superconductivity if $\lambda/\mu^{3-\Delta}\geq 2$ with $k/\mu=1/\sqrt{2}$, as illustrated in Fig.\ref{solu1}.
\end{itemize}
To see more details on the dependence of the critical temperature
on the lattice parameters, we may plot a 3D phase diagram on the
$k-\lambda$ plane. An example is shown in the middle plot of
Fig.\ref{solu2}, where we have set $e=2$. From this figure one can
obviously see that the phase transition occurs more easily in the
region with small lattice amplitudes but large wavenumbers, which
corresponds to the metallic phase before the transition (as is
shown in the left plot of Fig.\ref{solu2}). For a given
wavenumber, the condensate becomes harder with the increase of the
lattice amplitude, as illustrated in the right plot of
Fig.\ref{solu2}. While for a given lattice amplitude, the
condensate becomes harder with the decrease of the wavenumber (or
with the increase of the lattice constant).

\begin{figure}
\center{
\includegraphics[scale=0.85]{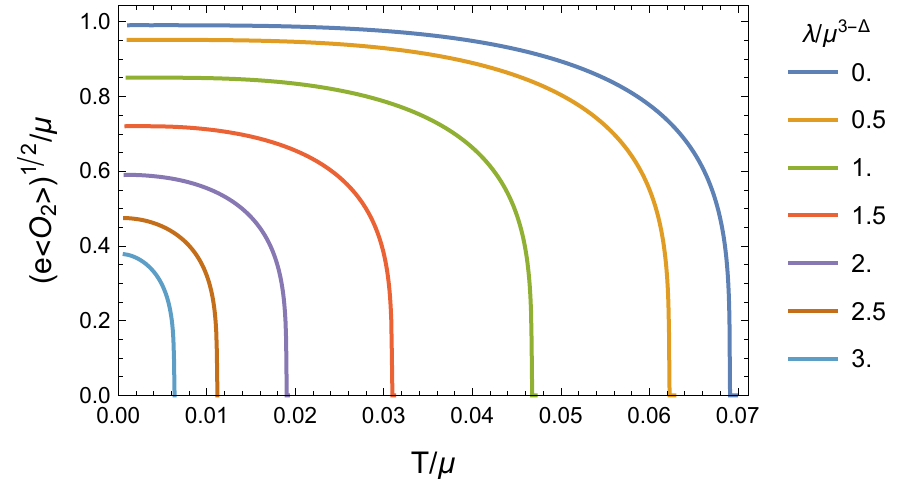}\hspace{0.2cm}
\includegraphics[scale=0.85]{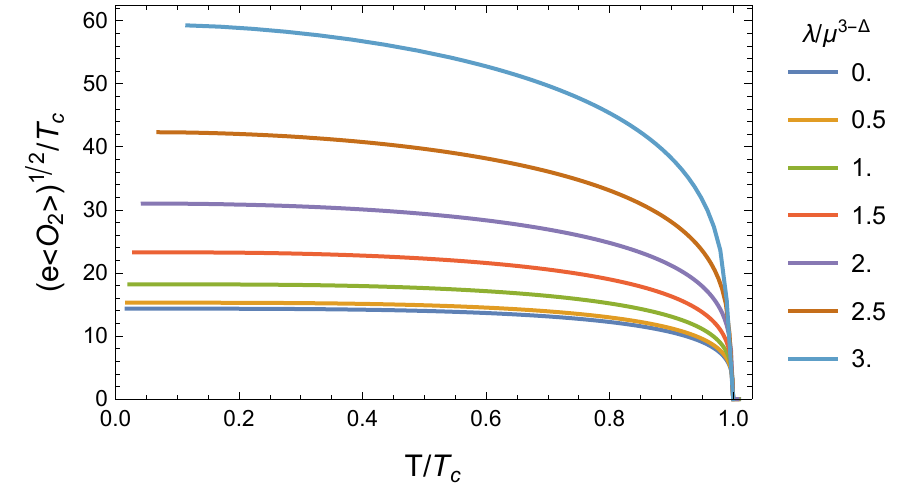}\hspace{0.2cm}
\caption{\label{solu3}The value of the condensate as a function of
temperature for various values of the lattice amplitude. The left
plot is shown in the unit of the chemical potential while the
right illustrates the same data in the unit of the critical
temperature. We have set $e=2$ and $k/\mu=1/\sqrt{2}$.} }
\end{figure}

Having found the critical temperature in a perturbative way, next
we will solve all the coupled equations of motion in
Eq.(\ref{eqom}) to find Q-lattice solutions with a scalar hair at
$T<T_c$. It involves in six ordinary differential equations with
variables $U,V_1,V_2,\psi,\phi$ and $\eta$, which can be
numerically solved with the standard pseudo-spectral method and
Newton iteration. We plot the value of the condensate as a
function of the temperature in Fig.\ref{solu3}.  From this figure it is
obvious to see that the critical temperature for the condensation
goes down when the lattice amplitude increases. Such a tendency
also implies that the condensation would never occur when the
amplitude is large enough and beyond some critical value.
Moreover, from the right plot in Fig.\ref{solu3} we find the expectation
value of the condensate becomes much larger in the unit of the
critical temperature, implying a larger energy gap $\omega_g/T_c$ for the
superconductor, {which seems also different from the results in
ionic lattices \cite{Horowitz:2013jaa}.} Finally, one can also fit
the data around $T=T_c$ and find the expectation value behaves as
$(1-T/T_c)^{1/2}$, indicating it is the second order phase
transition.

\begin{figure}
\center{
\includegraphics[scale=0.85]{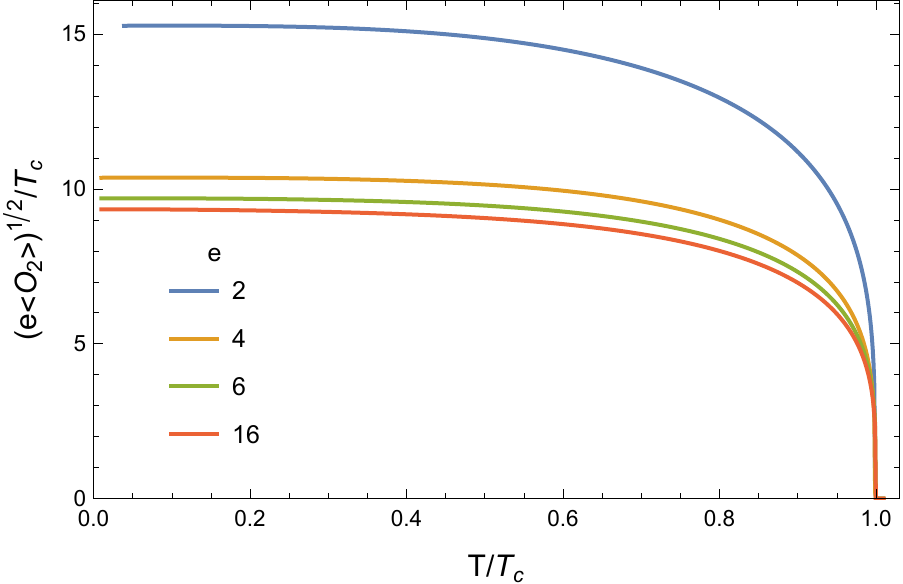}\hspace{0.5cm}
\includegraphics[scale=0.85]{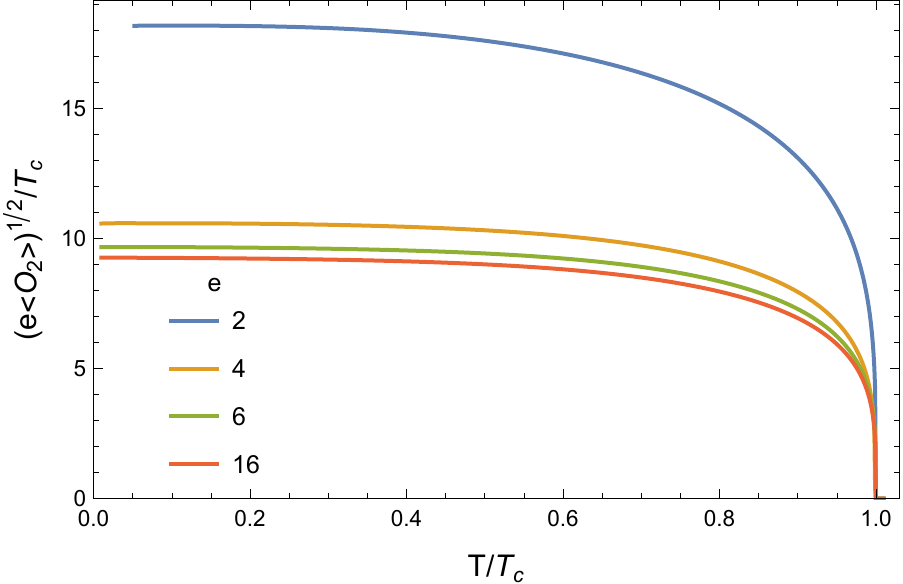}\hspace{0.5cm}
\caption{\label{solu4}The value of the condensate as a function of
the charge of the scalar field. Left plot we have set
$\lambda/\mu^{3-\Delta}=1/2$ and $k/\mu=1/\sqrt{2}$, while right
plot $\lambda/\mu^{3-\Delta}=3/2$ and $k/\mu=1/\sqrt{2}$.} }
\end{figure}

In the end of this section we address the issue of
 the relation
between the value of the condensate and the charge of the scalar
field. It is known in literature that when the back-reaction is
taken into account, the expectation value of the condensate which
may be denoted as $\omega_g$ depends on the charge of the scalar
field\cite{Hartnoll:2008vx}. But when $e\geq e_c\simeq
3\sqrt{2}$,\footnote{It corresponds to $e\geq e_c\simeq 3$ in
 the convention adopted in \cite{Hartnoll:2008kx,Hartnoll:2008vx}.} it is found that the
condensate will remain close to $\omega_g\sim 8\tilde{T}_c$. Later
this universal relation has been testified in various Einstein
gravity models with translational invariance. The presence of the
ionic lattice drops the critical charge down to a lower value and
it is found that even with $e\simeq 2$, a gap with $8T_c$ can be
reached \cite{Horowitz:2013jaa}. Now for Q-lattice background, we
plot the condensate as a function of the temperature for different
values of the charge in Fig.\ref{solu4}, where we have fixed
$\lambda/\mu^{3-\Delta}=1/2,3/2$, respectively and
$k/\mu=1/\sqrt{2}$. One can see that the condensate will approach
to $\omega_g\sim 9 T_c $ when the charge $e\geq e_c\simeq
6$.\footnote{We remark that the apparent discrepancy between our
result $\omega_g\sim 9 T_c $ and the well-known result
$\omega_g\sim 8 \tilde{T}_c $ in literature comes from the fact
that we have fixed the chemical potential and used it as the unit
of the system, while in previous literature one has a fixed charge
density and takes its square root as the unit. Thus $T_c$ and
$\tilde{T}_c$ have different units. We have checked that our
results are consistent with those in literature indeed once we
change the units.} Firstly, we have tested that this value, as
found in literature, is still universal in the probe limit (namely
$e\rightarrow \infty $) in our Q-lattice model and independent of
the values of lattice parameters. Secondly, in comparison with the
previous models we find the critical value of the charge becomes
larger. As a matter of fact, for different lattice parameters
$\lambda$ and $k$, we have a different value for the charge $e_c$
to characterize the critical region when $\omega_g\sim 9 T_c $.
Qualitatively, we find the larger the lattice amplitude is, the
larger the critical charge $e_c$. This tendency is consistent with
the fact that the presence of Q-lattice suppresses the condensate
of the scalar field.

\section{The optical conductivity }

Now we turn to compute the optical conductivity in the direction
of lattice. It turns out that it is enough to consider the
following consistent linear perturbation over the Q-lattice
background
\begin{equation}
\delta g_{tx}=h_{tx}(t,z), \delta A_{x}= a_{x}(t,z), \delta \Phi=
ie^{ikx}z^{3-\Delta}\varphi(t,z).
\end{equation}
As stressed in \cite{Donos:2013eha}, $h_{tx},a_{x}$ and $\varphi$
are {\it real} functions of {$(t,z)$} such that the perturbation
equations of motion will be real partial differential equations.
Moreover, we suppose the fluctuations of all the fields have a
time dependent form as $e^{-i\omega t}$. Thus again we are led to
three ordinary differential equations for $h_{tx}(z),a_{x}(z)$ and
$\varphi(z)$. Before solving these equations, we also mention that
besides the boundary condition $a_{x}(0)=1$, one more boundary
condition imposed at infinity is
$\varphi(0)=(ik\lambda/\omega)h^{(-2)}_{tx} $ where
$h^{(-2)}_{tx}$ is the coefficient of the leading order for metric
expansion at infinity $h_{tx}=h^{(-2)}_{tx}/z^2+\cdots$. Such a
boundary condition is to guarantee what we extract on the boundary
for the dual field is just the current-current correlator, as
investigated in \cite{Donos:2013eha} \footnote{ Such a boundary
condition is obtained by requiring that the non-zero quantities of
$h_{tx}$ and $\varphi$ on the boundary can be cancelled out by the
diffeomorphism transformation generated by the vector field
$\varsigma^x=\epsilon e^{-i\omega t}$, where $\epsilon$ is a small
parameter. Since here the scalar field of the background $\eta$ is
x-independent and invariant under this sort of diffeomorphism
transformation, this boundary condition remains in the
superconducting case.}. The optical conductivity is given by
\begin{equation}
\sigma(\omega)=\frac{\partial_z a_{x}(0)}{i\omega}.
\end{equation}

\subsection{Superconductivity over Q-lattices dual to a metallic phase}
\begin{figure}
\center{
\includegraphics[scale=0.75]{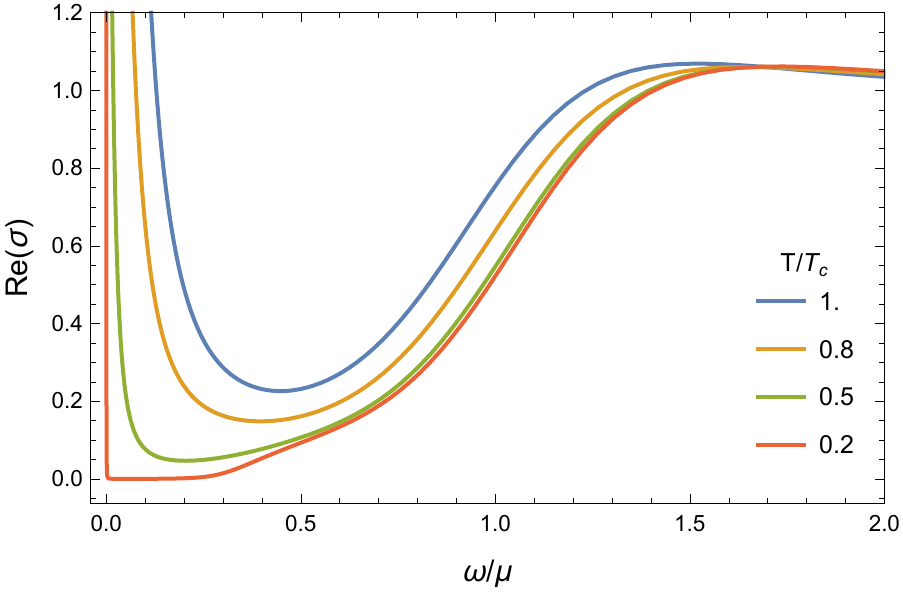}\hspace{0.5cm}
\includegraphics[scale=0.75]{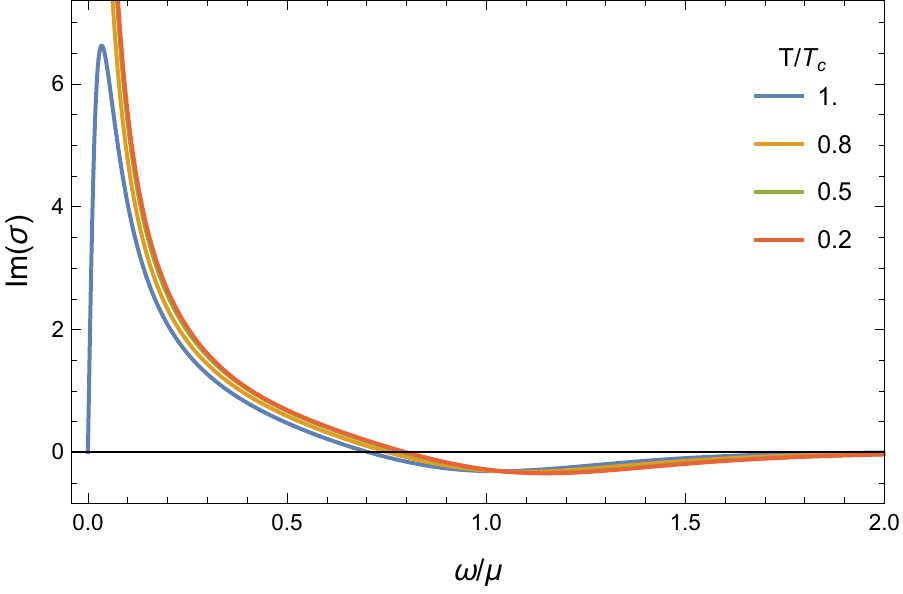}\hspace{0.1cm}
\includegraphics[scale=0.75]{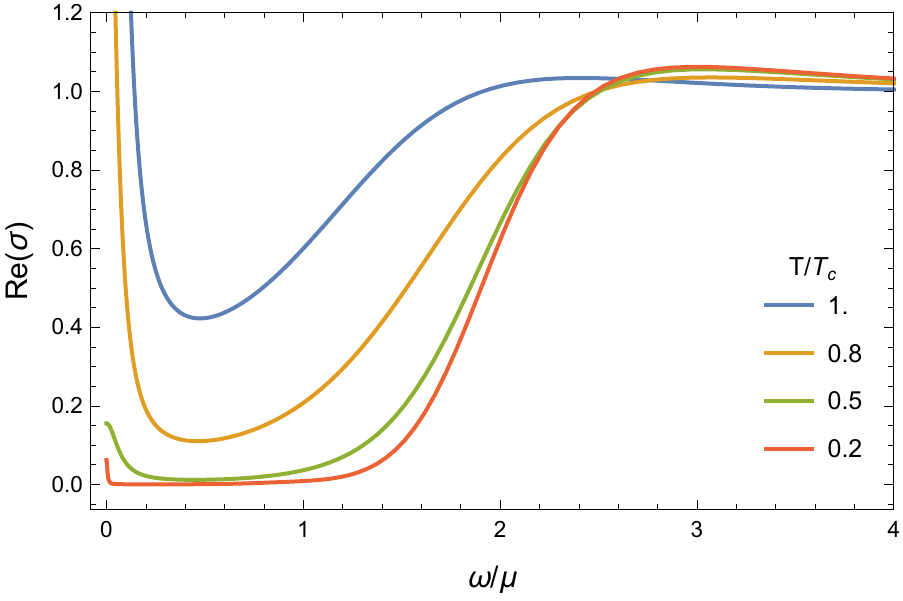}\hspace{0.5cm}
\includegraphics[scale=0.75]{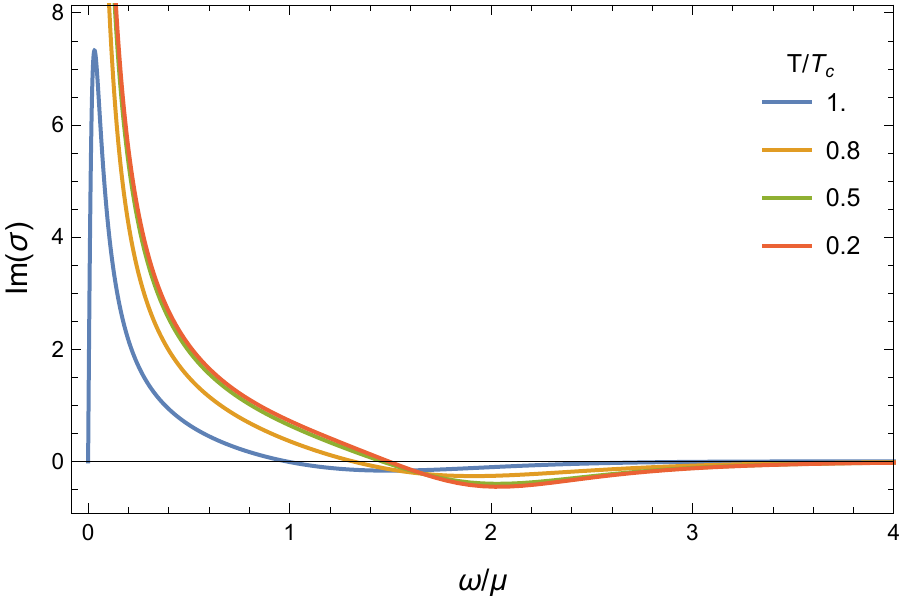}\hspace{0.1cm}
\includegraphics[scale=0.75]{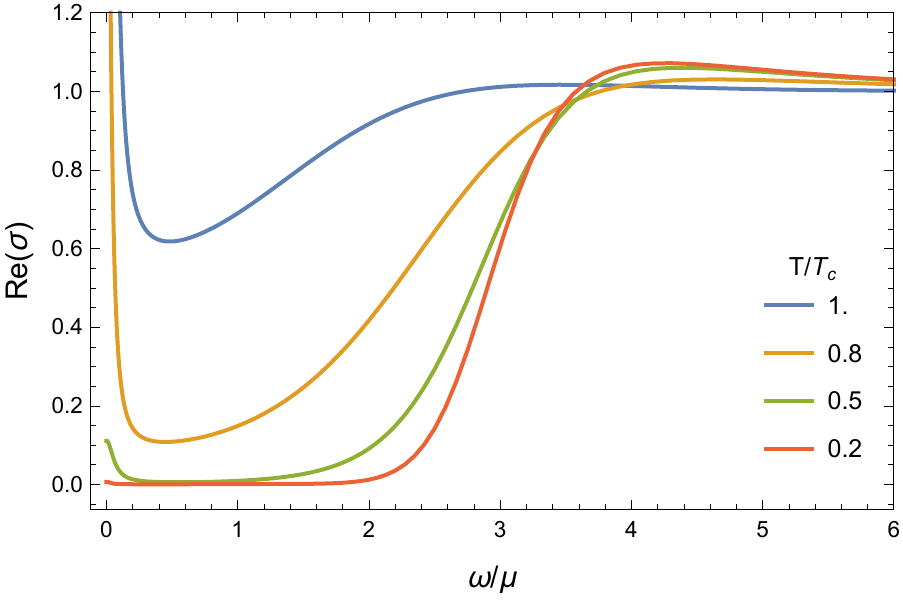}\hspace{0.5cm}
\includegraphics[scale=0.75]{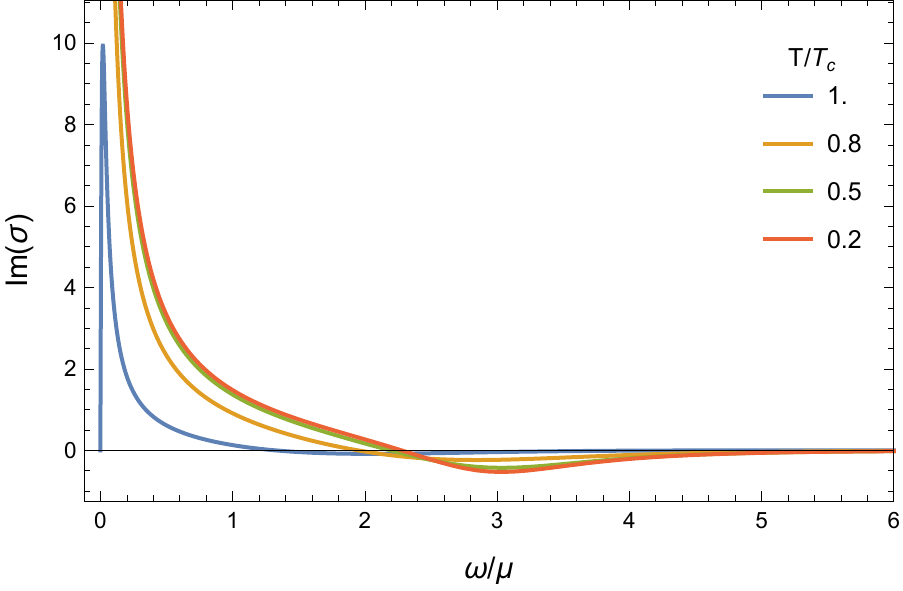}\hspace{0.1cm}
\caption{\label{cond1}The real and imaginary parts of optical
conductivity as a function of the frequency with
$\lambda/\mu^{3-\Delta}=1/2$ and $k/\mu=1/\sqrt{2}$. From top to
bottom the plots are for $e=2,4$,and $6$.} }
\end{figure}

In this subsection we discuss the optical conductivity as a
function of the frequency over a Q-lattice background which is
dual to a metallic phase before the phase transition. For
explicitness, we fix the lattice parameters as
$\lambda/\mu^{3-\Delta}=1/2$ and $k/\mu=1/\sqrt{2}$ in this
subsection. We show the real and imaginary parts of the
conductivity as a function of frequency for various charges in
Fig.\ref{cond1}. Our remarks on the behavior of the optical conductivity as
a function of frequency can be listed as follows.

\begin{figure}
\center{
\includegraphics[scale=0.58]{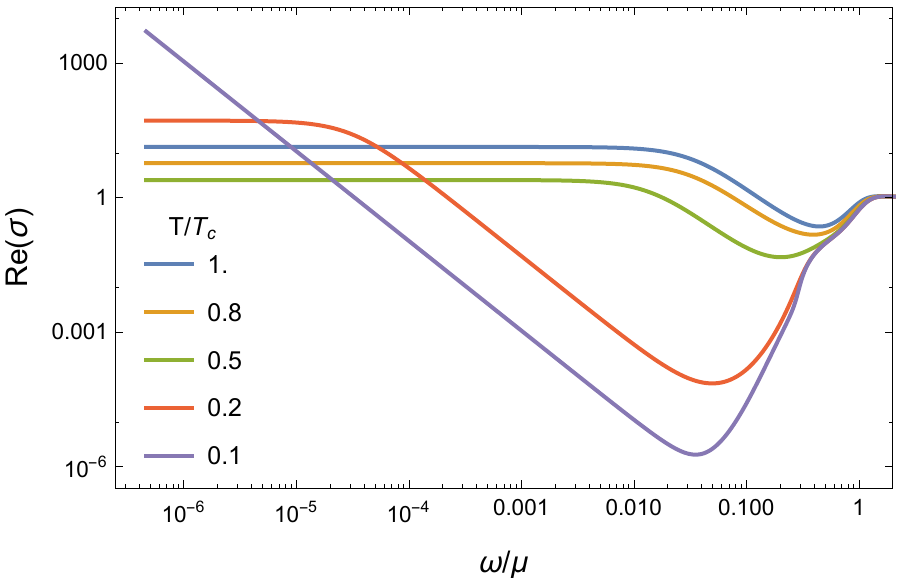}\hspace{0.1cm}
\includegraphics[scale=0.58]{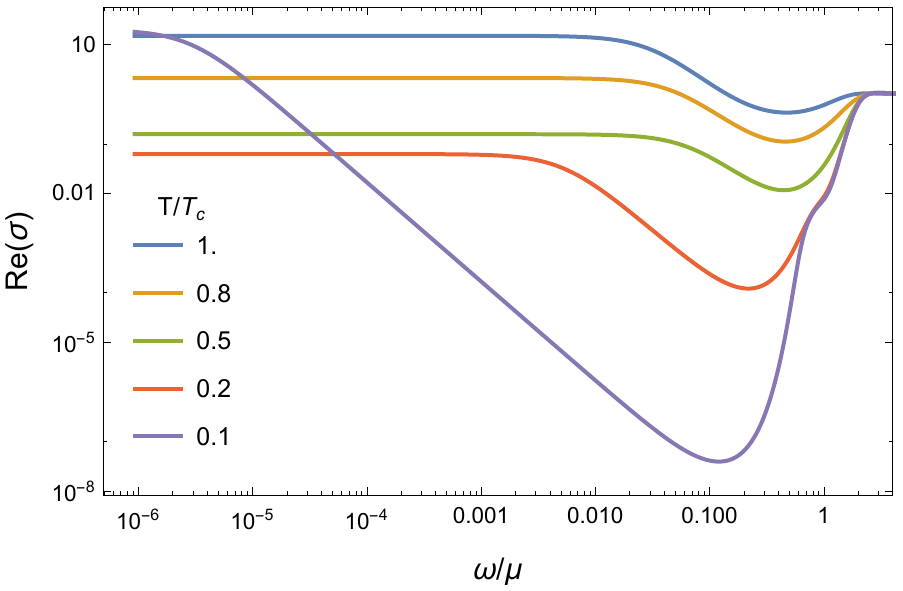}\hspace{0.1cm}
\includegraphics[scale=0.58]{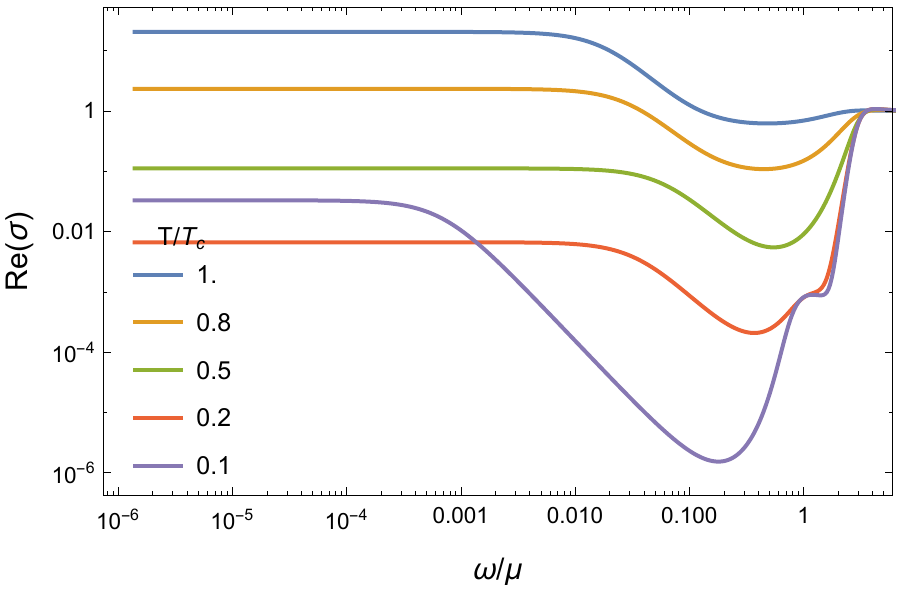}\hspace{0.1cm}
\caption{\label{log}The Log-Log plot of the real parts of optical
conductivity for Q-lattices with parameters given in Fig.\ref{cond1}. From
left to right the charge of the scalar field is taken as
$e=2,4$,and $6$.} }
\end{figure}

\begin{figure}
\center{
\includegraphics[scale=0.75]{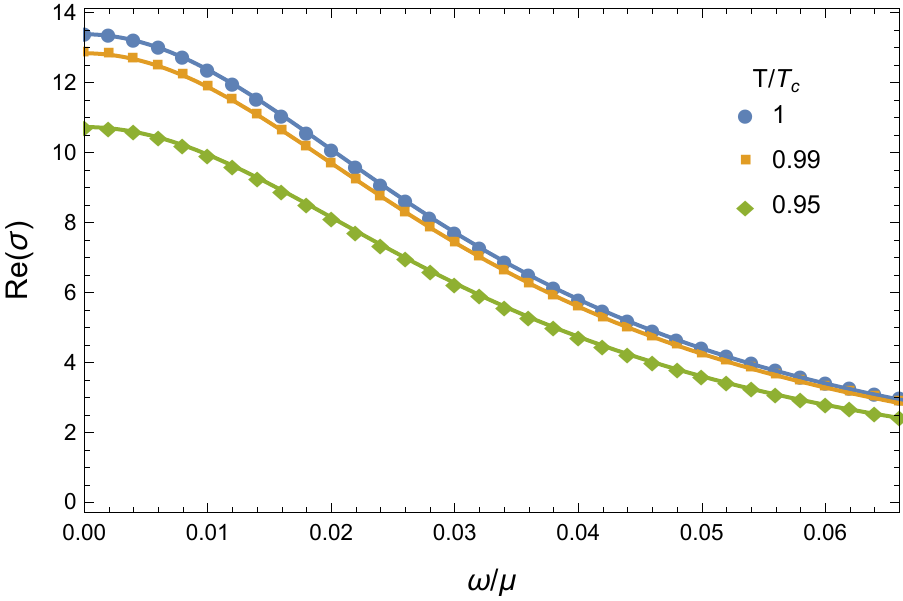}\hspace{0.1cm}
\includegraphics[scale=0.75]{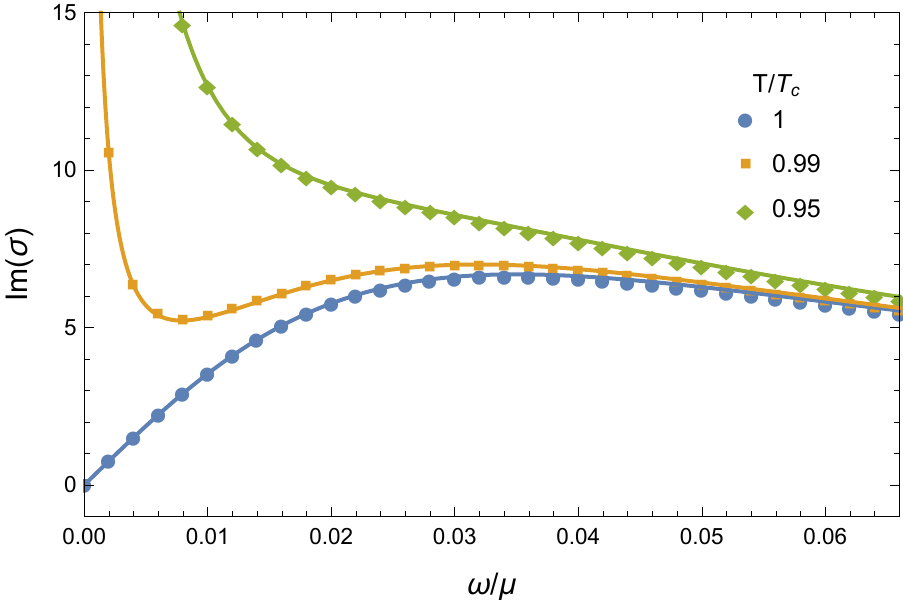}\hspace{0.1cm}
\caption{\label{drude}The critical behavior of the optical
conductivity near the critical temperature. The solid lines are
fits to Eq.(\ref{cd}).  } }
\end{figure}

\begin{figure}
\center{
\includegraphics[scale=0.57]{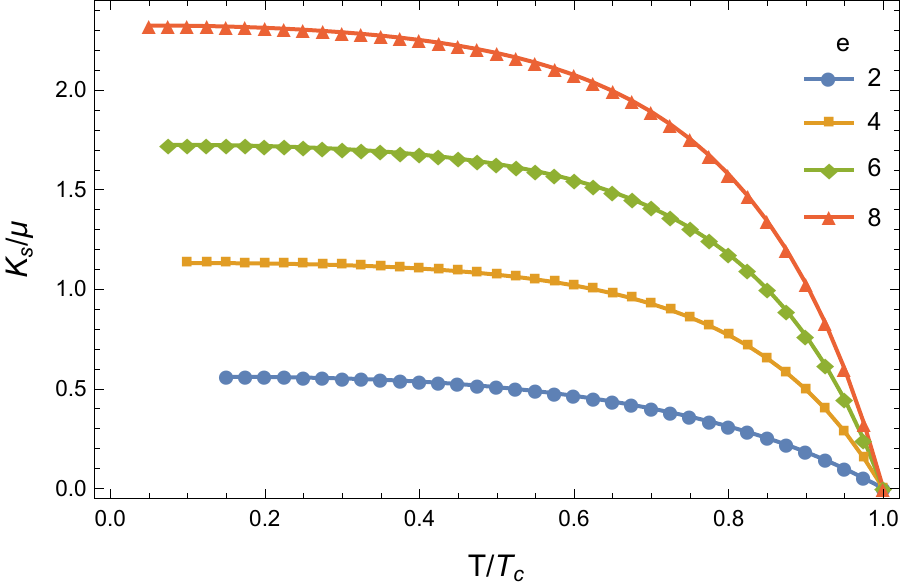}\hspace{0.1cm}
\includegraphics[scale=0.57]{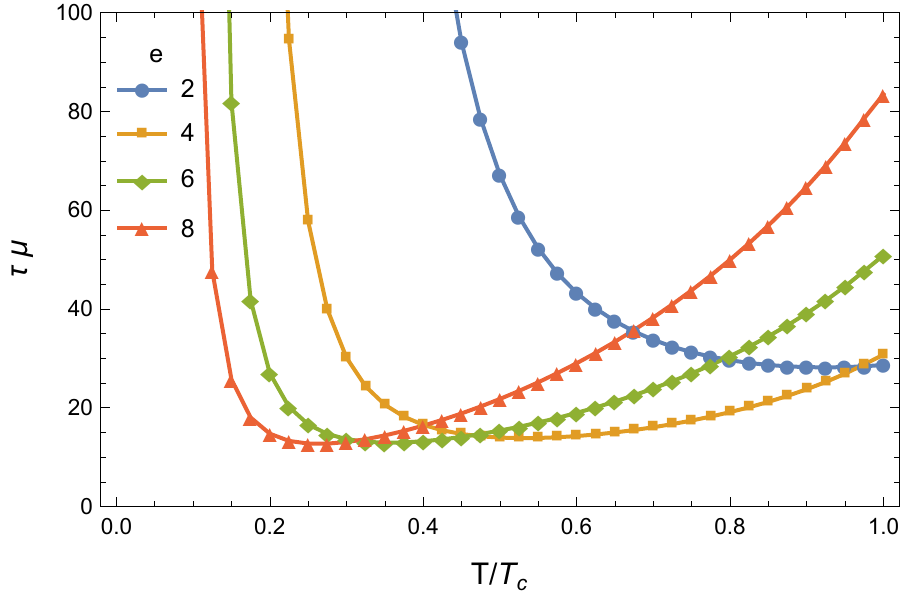}\hspace{0.1cm}
\includegraphics[scale=0.57]{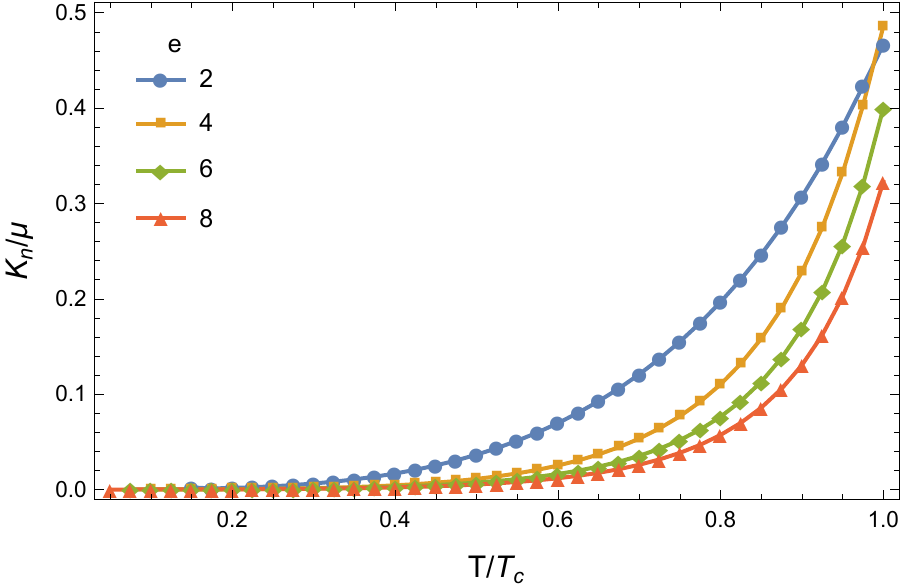}\hspace{0.1cm}
\caption{\label{density1} The three plots are
$K_s/\mu,K_n/\mu$ and $\tau\mu$ as functions of the temperature.} }
\end{figure}

\begin{itemize}
    \item {{\it Superconductivity}.}
First of all, in all plots we notice that {once the temperature falls
below the critical temperature, the imaginary part of the
conductivity will not be suppressed but climb up rapidly and
exhibit a pole at $\omega=0$.} Therefore, the Q-lattice does not
remove the delta function in the real part of the conductivity
below the critical temperature, confirming that it is dual to a
genuine superconductor.
    \item {{\it DC conductivity due to normal fluid}.}
The real part of the conductivity will rise at the low frequency
regime as well due to the lattice effects, indicating that there
is a normal component to the conductivity such that our
holographic model resembles a two-fluid model. Moreover, we notice
that the $DC$ conductivity
  will go down at first
 with the  decrease of the temperature and
 then rise up to a much larger value, which can become more transparent in
 a log-log plot as we show in Fig.\ref{log}. It means that normal
 component of the electron fluid is decreasing  to
 form  the superfluid component, but this normal component will not
 disappear quickly. The raise of the DC conductivity comes from the increase of the relaxation time, as we will describe below.
In addition, when the charge becomes larger, we find that the DC conductivity starts to rise up at  much lower $T/T_c$.
    \item {{\it Low frequency behavior}.}
In low frequency region we notice that the conductivity exhibits a
metallic behavior with a Drude peak even at much lower
temperature. We may fit the data at low frequency with the
following formula
\begin{equation}
\sigma(\omega)=\frac{iK_s}{\omega}+\frac{K_n\tau}{1-i\omega\tau},\label{cd}
\end{equation}
where $K_s$ and $K_n$ are supposed to be proportional to the
superfluid density $\rho_s$ and the normal fluid density $\rho_n$
, respectively, and $\tau$ is the relaxation time. We present a
fit to this equation near the critical temperature in Fig.\ref{drude} and
plot the values of these parameters as a function of the
temperature in Fig.\ref{density1}. In Fig.\ref{drude} the imaginary part of the
conductivity exhibits a sudden change in low frequency region when
the temperature drops through the critical point.  From Fig.\ref{density1} we
find that $K_s$ which is related to the superfluid density
increases as the temperature goes down and becomes saturated
around $T/T_c\simeq 0.6$, while $K_n$ which is related to the
normal fluid density decreases rapidly below the critical
temperature. However, the relaxation time does not have such a
monotonous behavior. As the temperature goes down from the
critical one, the relaxation time will decrease at first and then
rise up quickly in low temperature region, which looks peculiar in
comparison with other lattice models, where the relaxation time
monotonously increases with the decreasing of the temperature. In
particular, when the charge of the scalar field becomes large, the
turning point moves to lower temperature region. Such a phenomenon
might explain
 why we have a smaller DC conductivity at lower
temperature as described above, since it is proportional to the
relaxation time. But definitely, the issue of why the relaxation
time becomes smaller at lower temperature calls for further
understanding in the future. Finally, we remark that in log-log
plot the Drude behavior can also be conveniently captured by a
straight line with a constant slope as shown in Fig.\ref{log}, which has
previously been described in p-wave superconductors as well
\cite{Gubser:2008wv,Kuang:2011dy}.
\end{itemize}

Next we are concerned with the energy gap of the superconductor in
Q-lattice model, which may be evaluated by locating the minimal
value of the imaginary part of the conductivity at zero
temperature limit, which may be denoted as $(\omega/\mu)_{min}$.
For Q-lattices with parameters in Fig.\ref{cond1}, we find this value
corresponds to $(\omega/\mu)_{min}\simeq 18.454T_c,10.178T_c$, and
$9.247T_c$, respectively. Firstly, we find these values are
comparable with the values of the condensate we obtained in the
previous section and indeed they are close. Secondly, for
Q-lattices with lattice amplitude $\lambda/\mu^{3-\Delta}\sim
1/2$, $(\omega/\mu)_{min}$ will be saturated around $9T_c$ when
$e>e_c\simeq 6$. Finally, we have checked that the value
$(\omega/\mu)_{min}\simeq 9T_c$ is universal in the probe limit
$e\rightarrow \infty$, irrespective of the lattice parameters. In
general, the energy gap in zero temperature limit does depend on
the lattice parameters as well as the charge of the scalar field.
However, for a given Q-lattice with fixed $\lambda/\mu^{3-\Delta}$ and $e$, we
find {that} there always exists a critical value $e_c$  for the charge
such that the energy gap approaches
 the universal value
$(\omega/\mu)_{min}\simeq 9 T_c$ when $e\geq e_c$, which is
consistent with our analysis in the previous section on the
condensate of the background.

\begin{figure}
\center{
\includegraphics[scale=0.85]{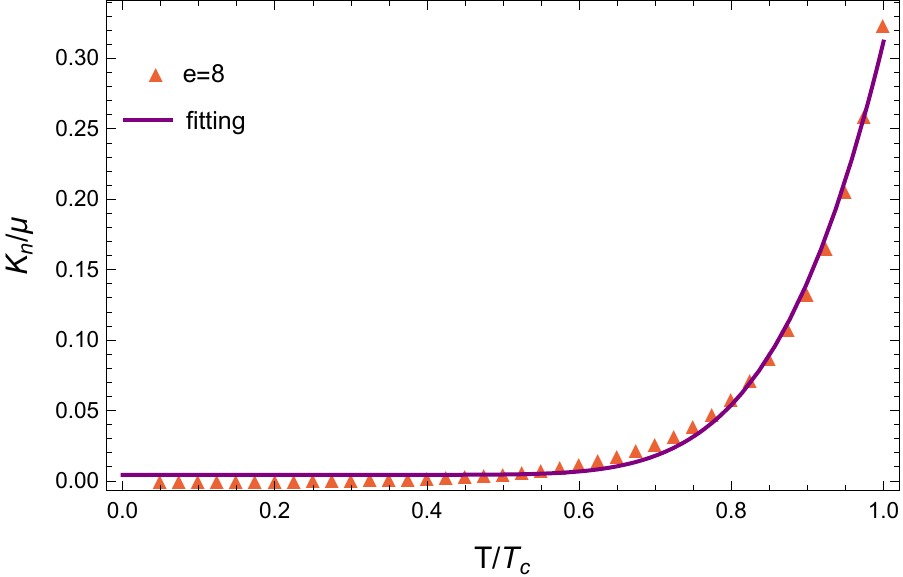}\hspace{0.1cm}
\includegraphics[scale=0.88]{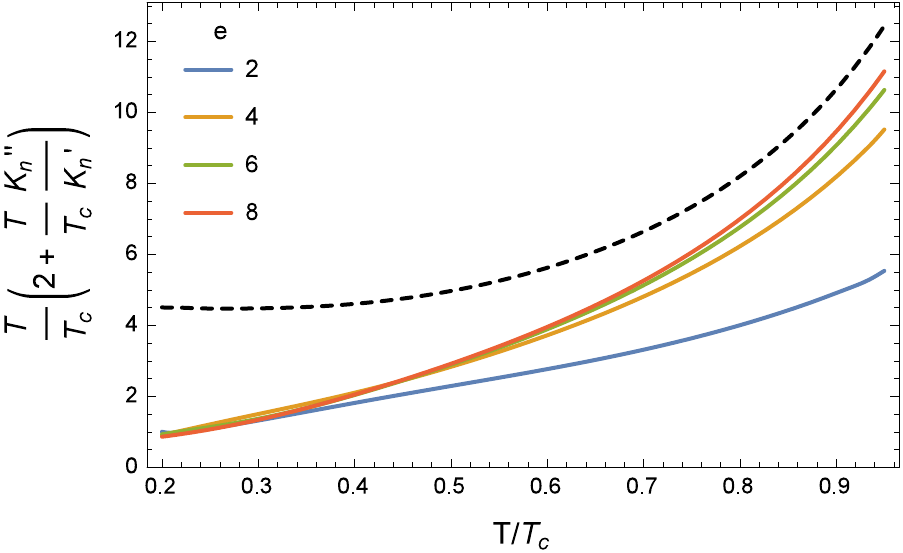}\hspace{0.1cm}
\caption{\label{density2} The first plot is the fit with Eq.(\ref{gapfit}). The second plot is for $\frac{T}{T_c}(2+\frac{T}{T_c}\frac{K_n''}{K_n'})$ versus $T/T_c$, where the derivative is with respect to $T/T_c$. The dashed black line in the last plot is for the holographic superconductor in the absence of the Q-lattice (Note that in this case the Drude law is absent, $K_n$ is identified with the $DC$ conductivity.). } }
\end{figure}

{In literature another way to evaluate the energy gap is to fit the temperature dependence of the normal fluid density in the zero temperature limit$(\Delta/T\gg 1)$ \cite{Horowitz:2013jaa}
\begin{equation}
\rho_n(T)=a + b e^{-\Delta/T}.\label{gapfit}
\end{equation}
In this thermodynamical method one need to know the normal fluid
density at first. Usually one assumes that $\rho_n\propto K_n$
such that the normal fluid density can be obtained by fitting the
conductivity with Eq.(\ref{cd}). Usually one expects that
$(\omega/\mu)_{min}\simeq 2 \Delta$ in the probe limit, which has
been testified in various holographic
models\cite{Hartnoll:2008vx,Hartnoll:2008kx,Horowitz:2013jaa}.
However, in the context of Q-lattice we find such exponential
behavior described by Eq.(\ref{gapfit}) is not clearly seen. The
left plot of Fig.9 is an attempt to fit $K_n/\mu$ with
Eq.(\ref{gapfit}), but obviously we notice that the data can not
be well fit in the entire region. To see if the data would have an
exponential behavior in any possible interval, we would better
take an alternative plot as follows. If the exponential
  behavior would present in some region, then from Eq.(\ref{gapfit}) one would
  find the quantity $\frac{T}{T_c}(2+\frac{T}{T_c}\frac{K_n''}{K_n'})$ should be a constant with value $\Delta/T_c$, irrespective of the value of
the parameter $a$, where the
  derivative is with respect to $T/T_c$. Therefore, an exponential behavior like Eq.(\ref{gapfit}) would be featured by
 a horizontal line in the plot of
$\frac{T}{T_c}(2+\frac{T}{T_c}\frac{K_n''}{K_n'})$ versus $T/T_c$.
Our results are shown in the second plot of Fig.9.  In Q-lattices
we do not find such behavior in any temperature region. The value
of $\frac{T}{T_c}(2+\frac{T}{T_c}\frac{K_n''}{K_n'})$ is not a
constant but varies with the temperature and is obviously below
the value for a system without Q-lattice in zero temperature
limit. In comparison, we notice that a holographic superconductor
without Q-lattice does exhibit such behavior in low temperature
region, which is shown as a dashed black line and points to
$\Delta/T_c\simeq 4.5$ as expected. Preliminarily we think this
discrepancy may imply that the factor $K_n$ might not be related
to the density $\rho_n$ simply by $K_n\propto \rho_n$ in Q-lattice
background. Recall that in theory $K_n=\rho_n e^2/m^*$, it would
be true when the effective mass of quasiparticles is also
temperature dependent. This issue deserves further study in the
future. In the end of this subsection we briefly address the issue
of the scaling law at the mid-frequency regime. This issue has
previously been investigated in both normal phase
\cite{Horowitz:2012ky,Ling:2013nxa,Donos:2013eha,Donos:2014yya,Bhattacharya:2014dea}
and superconducting phase\cite{Horowitz:2013jaa,Zeng:2014uoa}. It
was firstly noticed in the context of scalar lattices and ionic
lattices that in an intermediate frequency regime, the magnitude
of the conductivity exhibits a power law behavior as
\begin{eqnarray}
\label{PowerLaw} |\sigma(\omega)|=\frac{B}{\omega^{\gamma}}+C,
\end{eqnarray}
with $\gamma\simeq 2/3$ in four dimensional spacetime,
independent of the parameters of the model. This rule has been testified in
various models and in particular, its similarities with the
Cuprates in superconducting phase are disclosed. However, later it
is found that such a power law is not so robust in other lattice
models. In particular, it was pointed out that in the context of
Q-lattices there is no evidence for such an intermediate
scaling\cite{Donos:2013eha}. Now for the holographic
superconductors in Q-lattices, we may treat it in a parallel way
and our result is presented in Fig.\ref{scaling}. From this figure, we find
the intermediate scaling law is not manifest.
\begin{figure}
\center{
\includegraphics[scale=1]{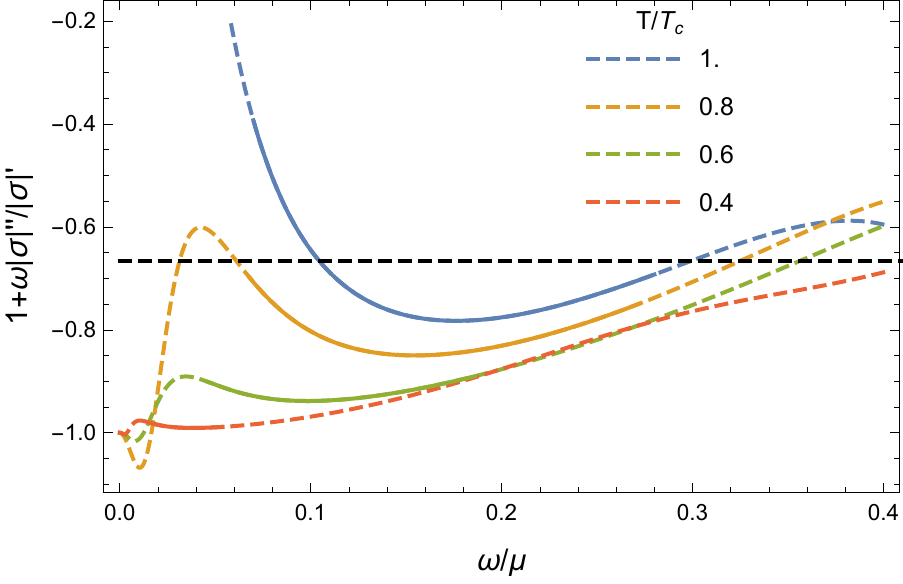}\hspace{0.5cm}
\caption{\label{scaling} The plot is for
$1+\omega|\sigma|^{''}/|\sigma|^{'}$ versus $\omega/\mu$. The
segment of each curve falling into the interval
$2\leq\omega\tau\leq 8$ is plotted as a solid line. } }
\end{figure}

\subsection{Superconductivity over Q-lattices dual to an insulating phase}

In this subsection we briefly discuss the superconductivity over a
Q-lattice which is dual to an insulating phase before the phase
transition\footnote{We mean the Q-lattice would exhibit an
insulating behavior in zero temperature limit in the absence of
the charged scalar field.}. We present a typical example with
$\lambda/\mu^{3-\Delta}=3/2$ and $k/\mu=1/\sqrt{2}$. The optical
conductivity for $e=4$ is plotted in Fig.\ref{cond2}. In comparison with
the superconductors over the Q-lattice in metallic phase, we
present some general remarks as follows.

\begin{itemize}
    \item With the decrease of the temperature, DC conductivity goes down at first and rises up again, which shares the temperature-dependence behavior with the case dual to the metallic phase.
    \item At low frequency, the lattice effects
drive the normal electron fluid to deviate from Drude relation and
exhibit an insulating behavior. This can be seen manifestly in the
log-log plot in which the straight line {with a constant slope
becomes shorter at lower temperatures}, as
illustrated in the last plot of Fig.\ref{cond2}.
    \item The energy gap has the same universal behavior in the probe limit, namely $(\omega/\mu)_{min}\simeq 9T_c$ as $e\rightarrow \infty$.
    For Q-lattices with parameters in Fig.\ref{cond2}, we have
$(\omega/\mu)_{min}\simeq 11.258T_c$, while for $e=6$, we have
$9.270T_c$.
\end{itemize}

\begin{figure}
\center{
\includegraphics[scale=0.58]{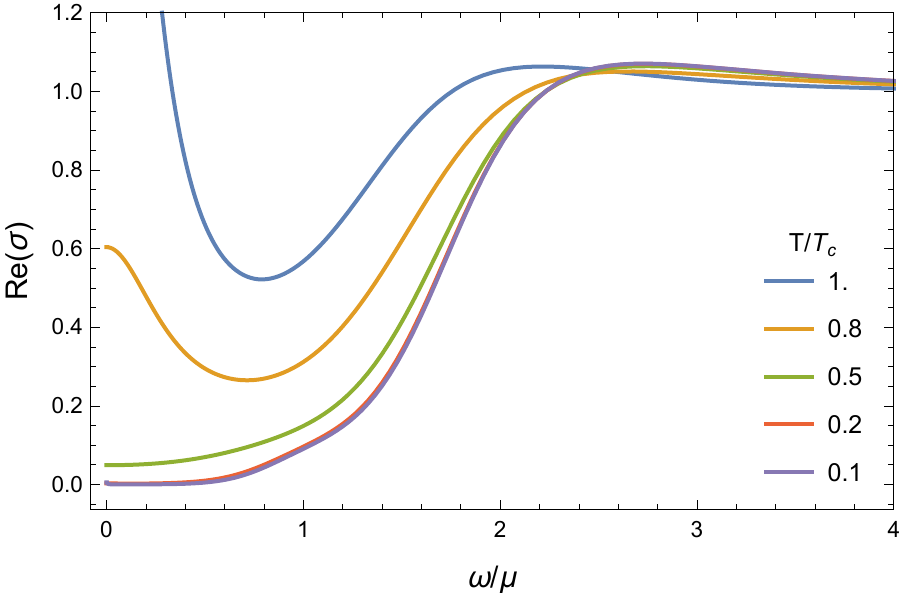}\hspace{0.1cm}
\includegraphics[scale=0.58]{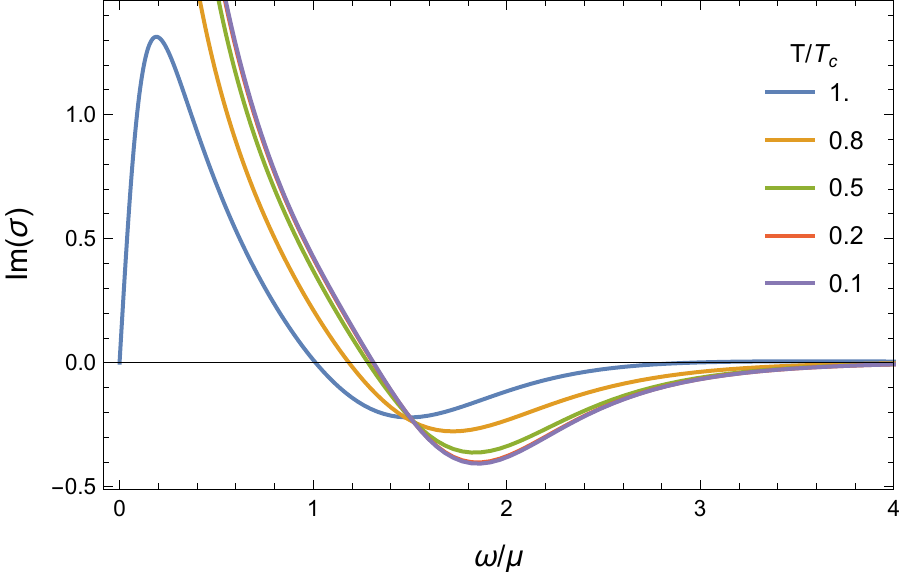}\hspace{0.1cm}
\includegraphics[scale=0.58]{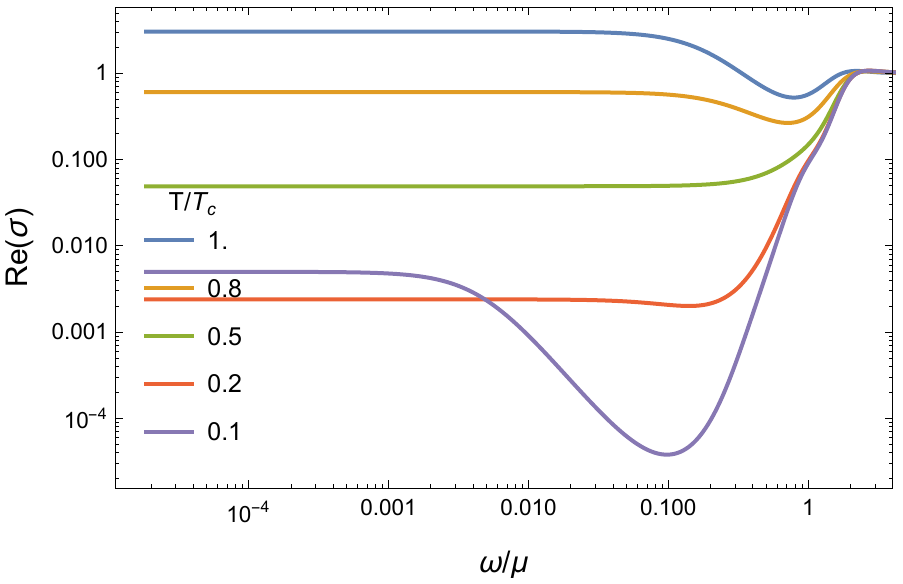}\hspace{0.1cm}
\caption{\label{cond2}The real and imaginary parts of optical
conductivity as a function of the frequency with
$\lambda/\mu^{3-\Delta}=3/2$, $k/\mu=1/\sqrt{2}$ and $e=4$. The
last one is a log-log plot for the real part.} }
\end{figure}

\section{Discussion}
In this paper we have constructed a holographic superconductor
model on Q-lattice background. We have found that the lattice
effects will suppress the condensate of the scalar field {and thus}
the critical temperature becomes lower in the presence of the
lattice. In particular, when the Q-lattice background is dual to a
deep insulating phase, the condensate would never occur when the
charge of the scalar field is relatively small. This is in
contrast to the results obtained in the context of ionic lattice
and striped phases, where the critical temperature is enhanced by
the lattice effects. In superconducting phase it is found that the
lattice does not remove the pole of the imaginary part of the
conductivity, implying the existence of a delta function in the
real part. The energy gap, however, depends on the lattice
parameters and the charge of the scalar field. Nevertheless, in
the probe limit, we find that gap $\omega_g\sim 9T_c$ is
universal, irrespective of the lattice parameters. This picture is
consistent with our knowledge on other sorts of holographic
superconductor models.

For convenience we have only computed the optical conductivity of
a superconductor where the Q-lattice background is dual to a
typical metallic phase or a typical insulating phase prior to the
condensation. In practice, many interesting phenomena have been
explored by condensed matter experiments in the critical region
where metal-insulator transition occurs in zero temperature limit.
From this point of view one probably shows more interests in the
superconducting behavior of the Q-lattice model with critical
parameters $\lambda_c$ and $k_c$. We leave this issue for
investigation in future.

This simplest model of holographic superconductors on Q-lattice
can be straightforwardly generalized to other cases. For instance,
we may consider to input Q-lattice structure in two spatial
directions with anisotropy \cite{Ling:2014bda}.
 We may also construct the
superconductor models on Q-lattice in other gravity  theories,
such as the Gauss-Bonnet gravity and Einstein-Maxwell-Dilaton
gravity. These works deserve
 further investigation.

\begin{acknowledgments}
This work is supported by the Natural Science Foundation of China
under Grant Nos.11275208, 11305018 and 11178002. Y.L. also
acknowledges the support from Jiangxi young scientists (JingGang
Star) program and 555 talent project of Jiangxi Province.
J. P. Wu is also supported by Program for Liaoning Excellent Talents in University (No. LJQ2014123).

\end{acknowledgments}

\end{document}